\documentclass[preprint,showpacs,preprintnumbers,amsmath,amssymb]{revtex4}
\usepackage{graphicx}
\begin{document}

\title{ Thermal smearing of the magneto-Kohn anomaly  for Dirac materials and comparison with the
two-dimensional electron liquid}

 \author{   Godfrey Gumbs$^{1,2}$,   Antonios Balassis$^3$, Dipendra Dahal$^1$,  M. L. Glasser$^{2,4}$ }
\affiliation{$^1$Department of Physics,
Hunter College of  City University of New York,
695 Park Avenue, New York, NY 10065, USA\\
$^2$Donostia International
Physics Center, P. Manuel de Lardizabal 4, \\20018 San
Sebastian, Spain \\
$^{3}$Department of Physics, Fordham  University, Bronx, NY 10458,  USA\\
$^{4}$Department of Physics, Clarkson University, Potsdam,
New York 13699-5820, USA
}

\date{\today}

\begin{abstract}

We compute  and compare the effects due to a uniform perpendicular  magnetic field as well as
 temperature  on the static polarization  functions  for monolayer graphene (MLG), associated
with the Dirac point, with that for   the two-dimensional electron liquid (2DEL)
with the use of  comprehensive numerical calculations.
The relevance of our study to the Kohn anomaly in low-dimensional structures  and the
 Friedel oscillations for the screening of the potential for a dilute distribution
 of impurities is reported. Our results show substantial differences due to
 screening for the 2DEL and MLG which have not
 been given adequate attention previously.
\end{abstract}

\pacs{ 73.21.-b,   71.10.-w,  73.43.Lp,  24.10.Cn, 68.65.Pq }

\maketitle

\section{Introduction}
\label{int}
There have been recent experimental reports regarding the observation
that there is  direct evidence  of
the role    played by screening  of   charged impurities in graphene and the two-dimensional
electron liquid on their properties \cite{a1,a2}.
Specifically, it has been demonstrated
that in the presence of a magnetic field the strength of the impurity scattering
 can be adjusted by controlling the occupation of Landau-level states,
which enters the polarizability.
The presence of impurities may produce potential fluctuations whose amplitude  can depend
on both the density of ions, their location, magnetic field and  temperature which can
affect their  high mobility.  These considerations are relevant in the design and
efficiency of semiconductor field effect transistors.
An interesting paper  in Ref.\ [\onlinecite{LaryG1}]  investigating the smearing of the Kohn anomaly
for a two-dimensional electron liquid (2DEL)  in a perpendicular weak magnetic field was published a
few years ago.  There,  the  principal goal was obtaining a closed-form analytic
  expression for the Landau level contribution to the  random-phase approximation (RPA)
static  polarization operator $\Pi(q , \omega=0)$, where $q$ denotes the wave 
vector transfer. Also, at that time,
a simple exact expression for $\Pi(q , \omega)$ valid for all
temperatures, frequencies $\omega$ and non-quantizing magnetic fields \cite{LaryG2}
existed in the literature (see also Refs. [\onlinecite{LaryG2+,LaryG2++}]) and
taking its derivative with respect to wave number may be carried out in a  straightforward way.

\medskip
\par
In this paper we intend to examine the behavior of the static polarizability and 
shielded potential in the strong field limit  for monolayer graphene and the 2DEL
 by carrying out detailed numerical calculations.
To  bolster confidence in our numerical procedure, we compare  our data
obtained with that already reported in the literature for special cases, as we have pointed out below.

\medskip
\par

We investigate the smearing of the Kohn anomaly in graphene
\cite{MAG1,MAG1+,MAG2,MAG3,MAG4,MAG5-1,MAG5,MAG5+,MAG6,IOP,pavlop,Screening-GR} and the 2DEL \cite{MAG1,MAG1+,GV}
by numerically calculating the static polarization function along with its derivative for
various magnetic field strengths and temperatures. We also report results for the screening
of an impurity embedded within or near a 2DEL and a graphene layer.  Our results at
zero-temperature are in general agreement with those presented in Ref.\ [\onlinecite{MAG1+}].

\medskip
\par

The outline of the rest of our presentation is as follows. In Sec.\ \ref{sec2},
we detail the Hamiltonian we employ for the Dirac electrons in MLG under
the influence of a perpendicular ambient magnetic field as well as
the 2D RPA ring diagram polarization function for graphene at finite temperature.
A careful calculation of the polarization function for a range of temperature
and magnetic field is reported in Sec.\ \ref{sec3} along with the associated screening
of impurities. We demonstrate that the analytic features in the polarization function are directly
reflected in those obtained for static shielding and so are detectable experimentally.
We conclude with highlights of our calculations in Sec.\ \ref{sec4}.

\section{Model for graphene layer and the Polarization Function in Magnetic Field}
\label{sec2}

Let us consider electrons in a single graphene layer in the $x-y$ plane
in a perpendicular magnetic field  $\mathbf{B}$ parallel to the
positive $z$ axis. The effective-mass Hamiltonian for noninteracting
electrons in  one valley in graphene in the absence of scatterers  is given
by the following equation. Here, we neglect the Zeeman splitting
and assume valley-energy degeneracy, describing the eigenstates
by two pseudospins. We have,
\begin{eqnarray}
\label{Ham}
\hat{H}_{0}= v_F \left(\begin{array}{cc}
0 &\hat{\pi}_{x} - i \hat{\pi}_{y} \\
\hat{\pi}_{x} + i \hat{\pi}_{y}&0
\end{array}   \right) \ ,
\end{eqnarray}
where $\hat{\mathbf{\pi}} = -i\hbar\nabla +e\mathbf{A}$, $-e$
is the electron charge, $\mathbf{A}$ is the vector potential,
 $v_F = \sqrt{3}at/(2\hbar)$ is the Fermi velocity with $a =2.57$ {\r A}
denoting the lattice constant and $t \approx 2.71$ eV is  the overlap integral
between  nearest-neighbor carbon atoms.

In the Landau gauge $\mathbf{A} = (0,Bx,0)$, the eigenfunctions
$\psi_\alpha(\mathbf{r})$ of the Hamiltonian $\hat{H}_{0}$ in Eq.\
(\ref{Ham})  are labeled by the set of quantum numbers
$\displaystyle{\alpha=\{k_y,n,s(n)\}}$ where $n=0,1, 2,\cdots$ is
the Landau level index, $k_y$ is the electron wave vector in the
$y$ direction, and $s(n)$ defined by
$ s(n) =  0$ for  $n=0$   and  $ s(n) =  \pm 1$ for  $n>0$
 indicates the conduction ($+1$) and valence ($-1$ and $0$) bands,
respectively. The two-component eigenfunction  $\psi_\alpha(\mathbf{r})$
is given  by
\begin{eqnarray}
\label{efunction}
\psi_{\alpha}(x,y) = \frac{C_{n}}{\sqrt{L_{y}}}e^{ik_y y}
\left( \begin{array}{c} s(n) i^{n-1}\Phi_{n-1}(x+l_B ^{2}k_y ) \\
i^{n}\Phi_{n}(x+l_B ^{2}k_y )
\end{array}\right)\ ,
\end{eqnarray}
where $l_B = \sqrt{\hbar /eB}$ is the magnetic length, and  $L_x$
and $L_y$ are normalization lengths in the $x$ and $y$ directions.
Also, $A=L_{x}L_{y}$ is the area of the system. We have
$C_n =   1$ for  $n=0$  and $1/\sqrt{2}$ for $n>0$.
 Additionally, we have
\begin{eqnarray}
\label{phin}
\Phi_n(x) = \left(2^{n}n!\sqrt{\pi}l_B \right)^{-1/2}
\exp\left[-\frac{1}{2}\left(\frac{x}{l_B }\right)^{2}\right]
H_{n}\left(\frac{x}{l_B }\right)\ ,
\end{eqnarray}
where $H_n(x)$ is a Hermite polynomial. The eigenenergies depend
 on the quantum numbers $n$ and $s$ only and are given by
\begin{eqnarray}
\label{energy}
\epsilon_{ns} = s(n) \epsilon_{n} = s(n) \frac{\hbar v_F }{l_B } \sqrt{2n}\ .
\end{eqnarray}
Note that the ratio of the Zeeman term $\Delta E_{Z}(B)$
to the  separation between  adjacent Landau levels $\Delta E_{L}(B) $
is negligible at high  magnetic field. For $B=10$ T, we have
$\Delta E_{Z}(B)/\Delta E_{L}(B) \approx \mu_{B}B/(\sqrt{2}\hbar v_F l_B ^{-1})
\approx 5\times 10^{-3}$. Here, $\mu_B = e\hbar /(2m_e)$ is the Bohr
magneton with $m_e$ denoting the free electron mass. Therefore, the
contributions to the single-electron Hamiltonian
from the Zeeman splitting and very small pseudospin splitting caused by
 two valleys in graphene may be neglected.
We assume
 energy degeneracy for the two possible spin projections
and two graphene valleys  described by pseudospins.

\medskip
\par

The 2D polarization function is given by \cite{MAG1,MAG1+,MAG2}

\begin{equation}
\Pi{(q,\omega)}=\frac{g}{2\pi l_B ^2}\sum_{n=0}^{\infty}
\sum_{n'=0}^{\infty}\sum_{s,s'=\pm} \frac{f_0(\epsilon_{n s})-f_0(\epsilon_{n's'})}{\epsilon_{ns}-\epsilon_{n's'}+\hbar\omega+i\delta}F_{nn'}^{ss'}(q) \ ,
\label{ref:5}
\end{equation}
where, $g=4$ for the spin and valley degeneracy, $f_0(\epsilon_{ns})
= 1/\left[e^{(\epsilon_{ns}-\mu)/k_BT}+1]\right]$ is the Fermi distribution function,with  $s$,$s'=\pm1$,
$\epsilon_{ns}=(s \hbar v_F/l_B)\sqrt{2n}-E_F$, where $E_F$ is the Fermi energy and the form factor
 $F_{nn'}^{ss'}(q)$ is defined as
\begin{eqnarray}
F_{nn'}^{ss'}(q)&=& e^{-l_B^2 q^2/2}\Bigg(\frac{l_B^2 q^2}{2}
\Bigg)^{(n_>-n_<)}
\nonumber\\
&\times& \Bigg(s 1_n^\ast 1_{n'}^\ast
\sqrt{\frac{(n_<-1)!}{(n_>-1)!}}L_{n_<-1}^{n_>-n_<}(l_B^2q^2/2)+s'2_n^\ast 2_{n'}^\ast
\sqrt{\frac{n_<!}{n_>!}}L_{n_<}^{n_>-n_<}(l_B^2q^2/2)\Bigg)^2
\end{eqnarray}
 Here,  $1_n^\ast=[(1-\delta_{n,0})/2]^{1/2}$, $2_n^\ast=[(1+\delta_{n,0})/2]^{1/2}$,
 $n_>$=$\max(n,n')$,
 $n_<$=$\min(n,n')$ and $L_n^a(x)$ is the associated Laguerre polynomial.
At $T=0$ K we have,
\begin{equation}
\Pi_{\rm inter}{(q,\omega)}=\sum_{n=0}^{N_c}\sum_{n'=N_F}^{N_c}\frac{F_{nn'}^{-+}}{-\epsilon_n-\epsilon_{n'}+\hbar\omega
+i\delta}+(\hbar\omega_+\rightarrow \hbar\omega_-)
\end{equation}
with, $N_F$ being the filling factor, $\hbar\omega_{\pm} =\pm (\hbar\omega+i\delta)$, $s = -1$ and $s^\prime= +1$. We also have
\begin{equation}
\Pi_{\rm intra}{(q,\omega)}=\sum_{n=0}^{N_F-1}\sum_{n'=N_F}^{N_c}
\frac{F_{nn'}^{++}}{\epsilon_n-\epsilon_{n'}+\hbar\omega +i\delta}+(\hbar\omega_+\rightarrow \hbar\omega_-)
\end{equation}
and  the total polarizability is obtained from
\begin{equation}
\Pi(q,\omega)=\Pi_{\rm inter}(q,\omega)+\Pi_{\rm intra}(q,\omega)\ ,
\end{equation}
where $\Pi_{\rm inter}(q,\omega)$ is due to transitions from the valence to
the conduction band while $\Pi_{\rm intra}(q,\omega)$ accounts for transitions within the
conduction band.
We  must  also note that the chemical potential depends on temperature approximately as

\begin{equation}
 \mu \approx  E_F  \left\{ 1  -  \frac{\pi^2}{6} \frac{d\ln(\rho(E_F ))}{d\ln(E_F )}  \left(
\frac{k_BT}{E_F }\right)^2 + \cdots  \right\}
\end{equation}
where $\rho(E_F )$ is the density-of-states at the Fermi energy $E_F $.
  Of course, for graphene, we  have $\rho(\epsilon)=\epsilon/[\pi(\hbar v_F)^2] $
  whereas for the 2DEL  the density of states is constant and given by
  $\rho(\epsilon)= m^\ast /\pi\hbar^2$ so that we can replace the chemical potential
  by the Fermi energy when the temperature of the 2DEL is low.
        Furthermore, in calculating the temperature-dependence of the polarization
        function at low temperature ($k_BT\ll E_F $), we employ
        $f_0(\epsilon;T)\approx  \theta(E_F -\epsilon)- (k_BT)\delta(\epsilon-E_F )$
        in terms of the Heaviside step-function $\theta(x)$.

        In evaluating the finite-temperature polarization, we may employ the following transformation
        of Maldague   \cite{LaryG2++} relating its values in the absence ($T=0$) and presence of a heat bath
        in the absence of magnetic field, i.e.,


\begin{equation}
        \Pi(q,\omega;T) = \int_0^\infty d\,E
        \frac{\Pi_{T=0,E_F=E  }(q,\omega  )}{4 k_BT
  \cosh ^2\left[ \frac{E-\mu(T)   }{2 k_B T}
\right]  }  \ .
\end{equation}
Recently, a generalization of this transformation to graphene when an external magnetic
field is applied was given in \cite{pavlop}.
At high temperatures or  weak magnetic field, when the separation
between Landau levels is small, and $k_BT \gg \hbar v_F /l_B $, the
occupation of the Landau levels is given by the Fermi-Dirac distribution
function $f_{n,s(n)} = 1/\left[e^{\left(\epsilon_{n,s(n)}/k_B T)\right)} + 1 \right]$
for $n>0$  if the energy is measured from the Fermi level.
When the magnetic field is sufficiently high, Landau level separation
is large and we may take only a few terms in the polarization sums since
transitions to higher levels have smaller oscillator strengths.
Furthermore, the separation between  Landau levels decreases as $n$
increases and is $\propto 1/\sqrt n$, for large $n$. In our calculations, we included Landau levels with $1<n\leq 70$ unless stated otherwise.
We may separate the contributions to the polarization  $\Pi(q ,\omega)$ into contributions
which  correspond to transitions between different Landau
levels inside the conduction band, and this term does not contribute
at $T=0$ K.

\medskip
\par

We may now make use of our result for the polarization function to calculate the shielded potential
of a point charge on the polar axis at a distance $z_0$ from the 2D plane. This is given as
a Hankel transform of order zero by \cite{Visscher}
\begin{equation}
V\left(  r_\parallel , z_0\right) = \left( \frac{ e^2}{\epsilon_s}  \right) \int_0^\infty  dq
\ \  J_0(q  r_\parallel)
e^{-q  z_0}  \frac{1}{\epsilon(q ,\omega=0)}
\label{VV}
\end{equation}
where the dielectric function is obtained in the random-phase-approximation (RPA)  from
\begin{equation}
\epsilon(q ,\omega)=1- \frac{2\pi e^2}{\epsilon_s q } \Pi  \left( q ,\omega  \right) \
\end{equation}
and $\epsilon_s=4\pi\varepsilon_0\epsilon_b$ with $\varepsilon_0$ the permeability of free space
and $\epsilon_b$ the background dielectric constant. As a matter of fact, Eq.\ (\ref{VV})
determines the static density distribution around an  impurity, as well as the effective interaction
between two test charges embedded within the 2D structure. In general, the screened potential
decreases exponentially with $z_0$ and is not of interest to us.

\vskip 0.3in

 \section{Numerical Results and Discussion}
 \label{sec3}

In our numerical calculations, we chose $r_s=3.0$ for the 2DEL where we define
$r_s=2m^\ast e^2/(\epsilon_b k_F\hbar^2)$ in terms of the electron effective
mass $m^\ast$, the background dielectric constant $\epsilon_b$ and
the Fermi wave vector $k_F$. We chose $m^\ast=0.067\ m_e$ ($m_e$ is the free electron mass) and $\epsilon_{b}=13.6$ which are appropriate for GaAs/AlGaAs. Then $r_s$ corresponds to electron density $n=6.1\times10^{10}\ {\rm cm}^{-2}$. For monolayer graphene, we chose  $r_s=1.0$ where
now we define $r_s$ by $r_s= e^2/(\epsilon_b \hbar v_F)$ given in terms of
the Fermi velocity $v_F$ for pristine graphene and the background dielectric
constant for this material.

\medskip
\par

\begin{figure}[t]
\centering
\includegraphics[width=3.67in,height=2.41in,keepaspectratio]{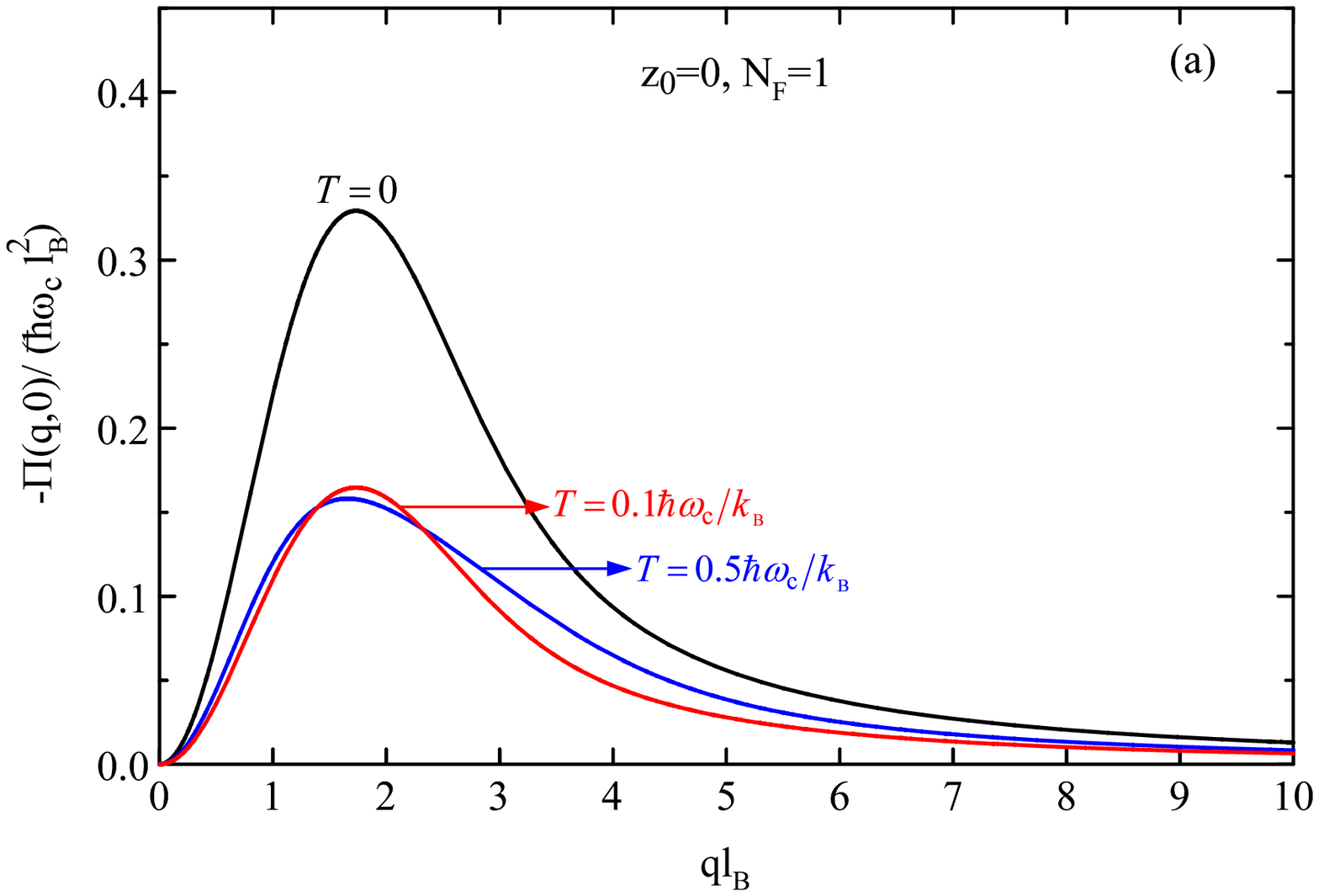}
\includegraphics[width=3.67in,height=2.41in,keepaspectratio]{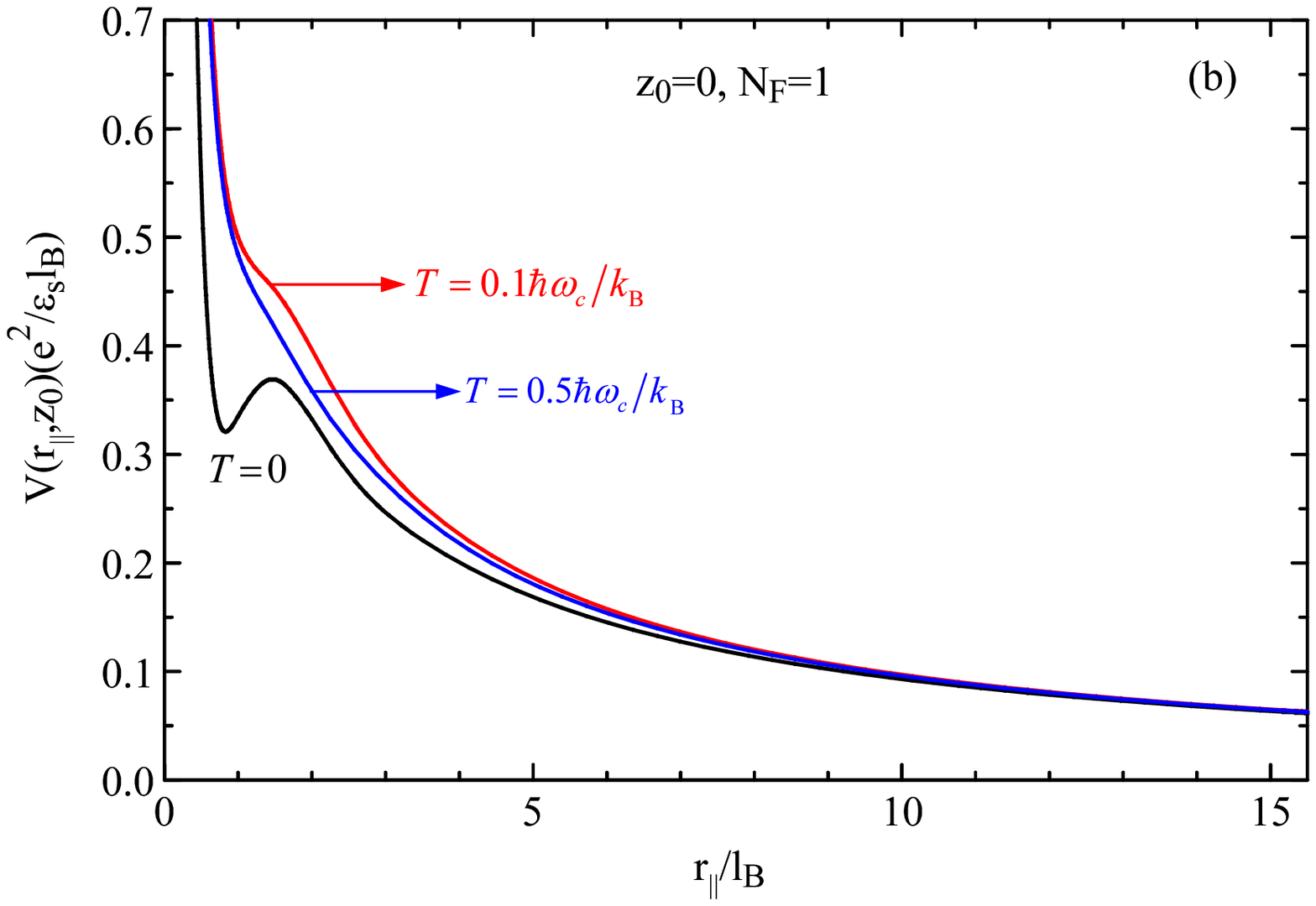}
\includegraphics[width=3.67in,height=2.41in,keepaspectratio]{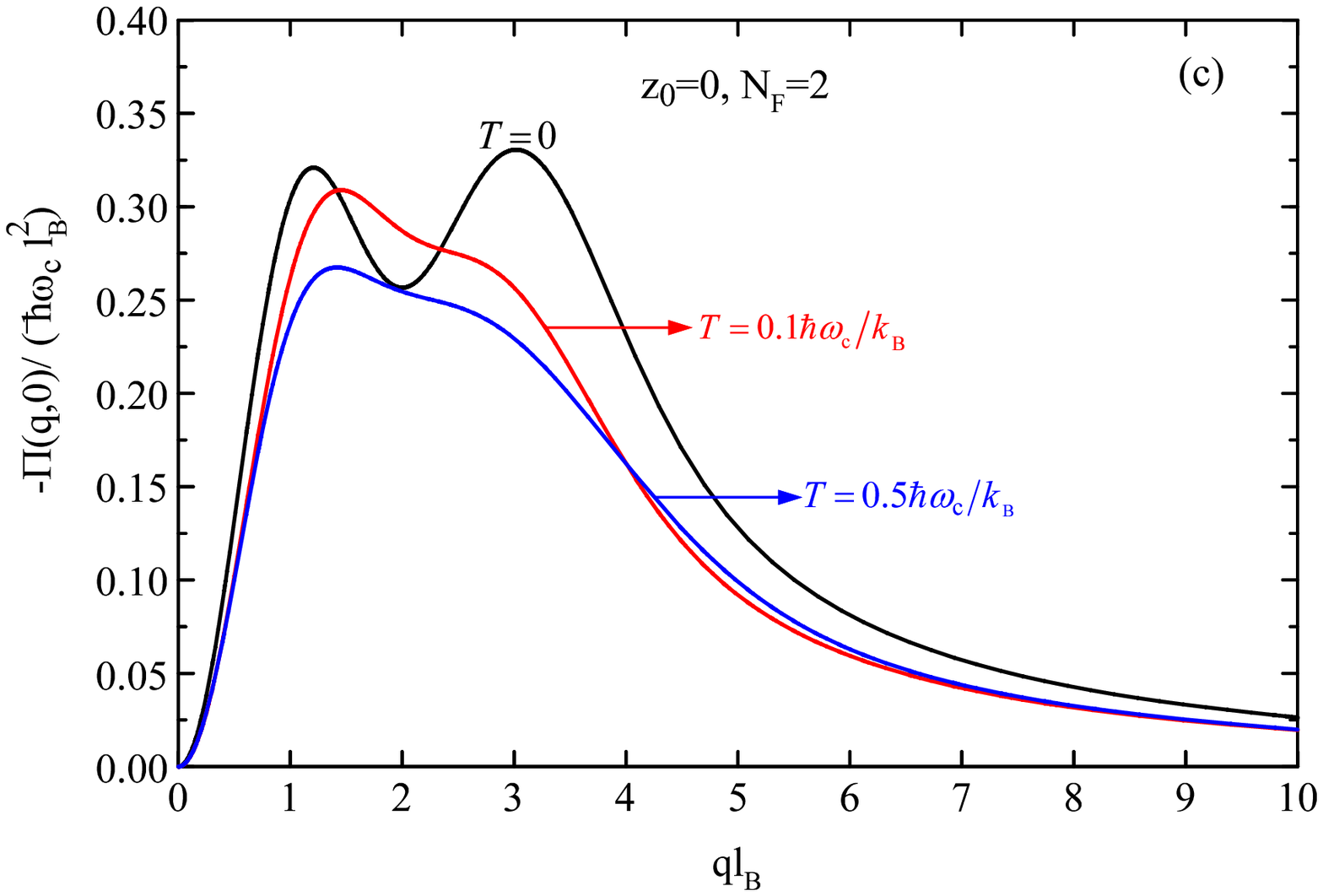}
\includegraphics[width=3.67in,height=2.41in,keepaspectratio]{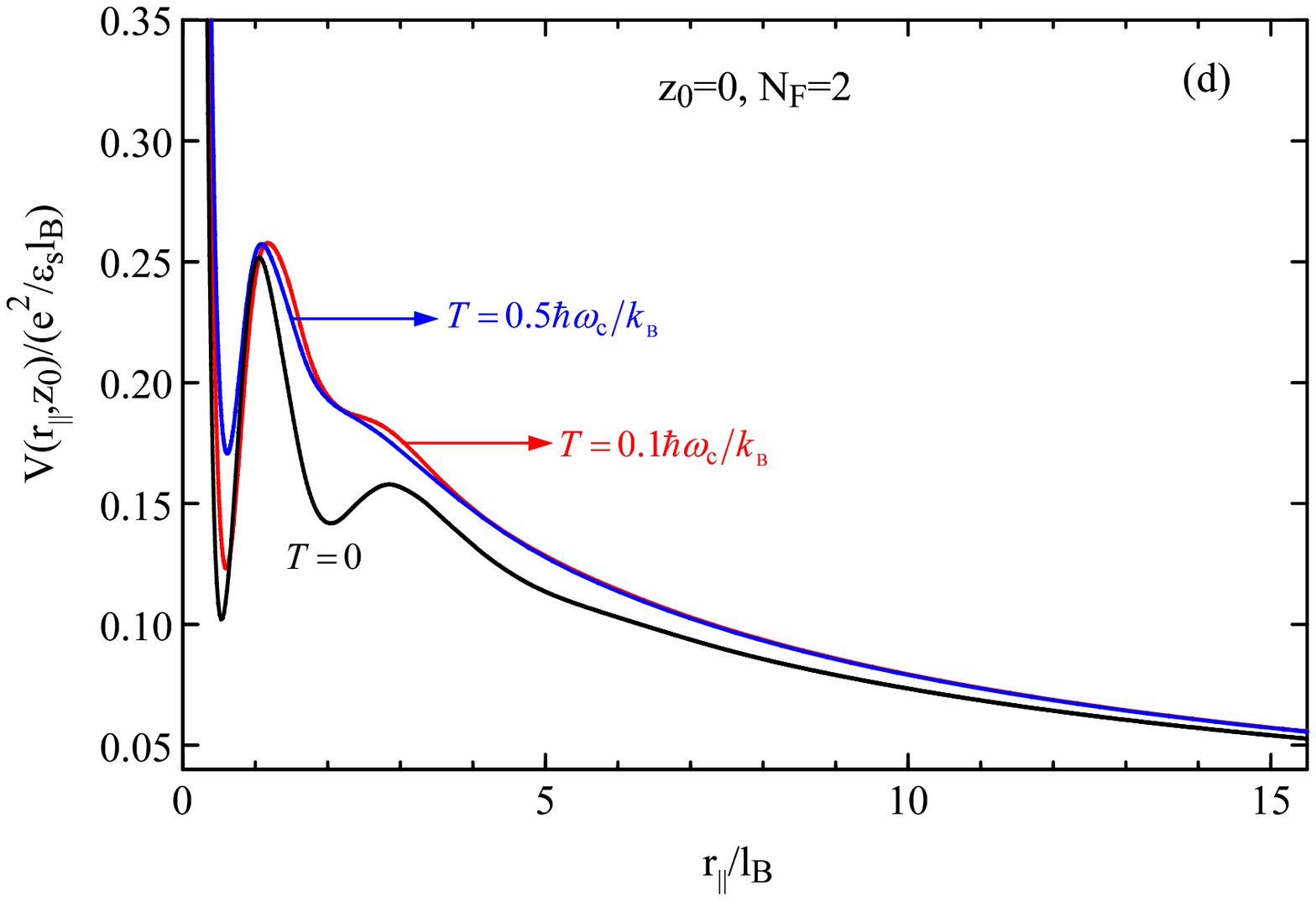}
\caption{(Color online) For two filling factors $N_F=1,2$, the static polarization function   (left panels)
for the 2DEL  for various   temperatures   as a
function of the wave vector $q   $ in units of the inverse magnetic
length $l_B^{-1}$.   The right panels show the corresponding screened  potentials.
 }
\label{FIG:1}
\end{figure}

In  Fig.\ \ref{FIG:1},    we present results from our calculations for the  static polarization function
$\Pi(q ,\omega=0)$  and the shielded potential   for a 2DEL  in the presence of a perpendicular
magnetic field  for two filling factors  $N_F$   at   chosen temperatures.
At $T=0$ K, the polarization function has
 $N_F$ sharp local maxima corresponding to the number of completely full Landau levels. As the
 temperature is increased, these peaks become smeared out and the polarizability is diminished.
 Correspondingly, the screened potential  exhibits  exactly the same number of local maxima as its
 polarization at $T=0$ K responsible for dielectric screening.  Just like the polarizability,
 there is smearing of these peaks when the temperature is increased
\cite{pavlop}.
We have $N_F=\pi l_B^2n_{2D}$  at $T=0$ K in terms of the electron density $n_{2D}$. The Fermi wave vector $k_F$ is also given by
$k_Fl_B=\sqrt{2N_F-1}$.
At  $T=0$ K, when there are multiple peaks, these peaks increase in height with increasing
wave vector    \cite{GV}. But, at finite temperature, the height
of the peaks decreases with increasing wave vector.
Irrespective of what temperature chosen, the first peak in the static polarizability
function  appears at $q \approx   2k_F$.

\medskip
\par

\begin{figure}[t]
\centering
\includegraphics[width=3.67in,height=2.41in,keepaspectratio]{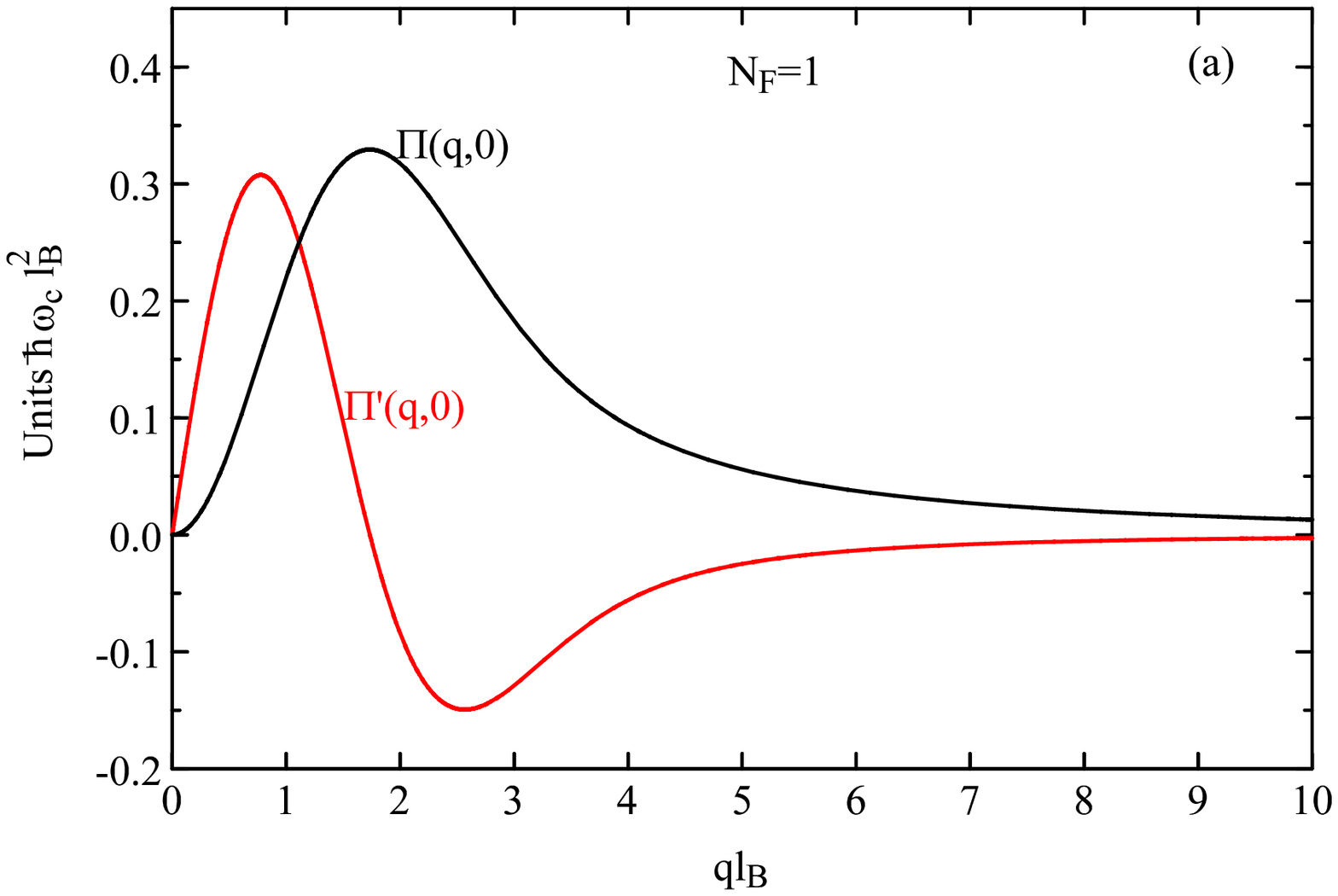}
\includegraphics[width=3.67in,height=2.41in,keepaspectratio]{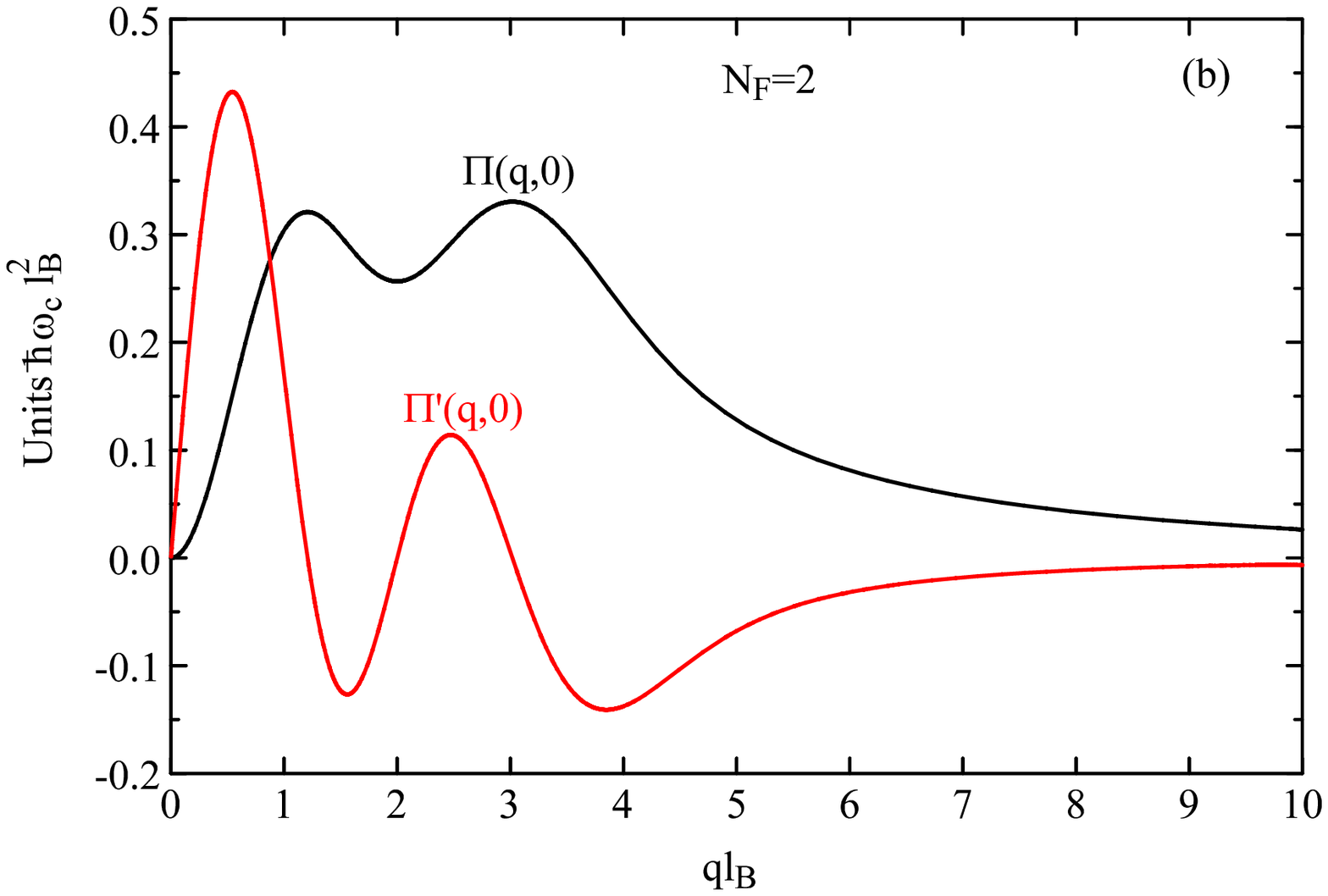}
\includegraphics[width=3.67in,height=2.41in,keepaspectratio]{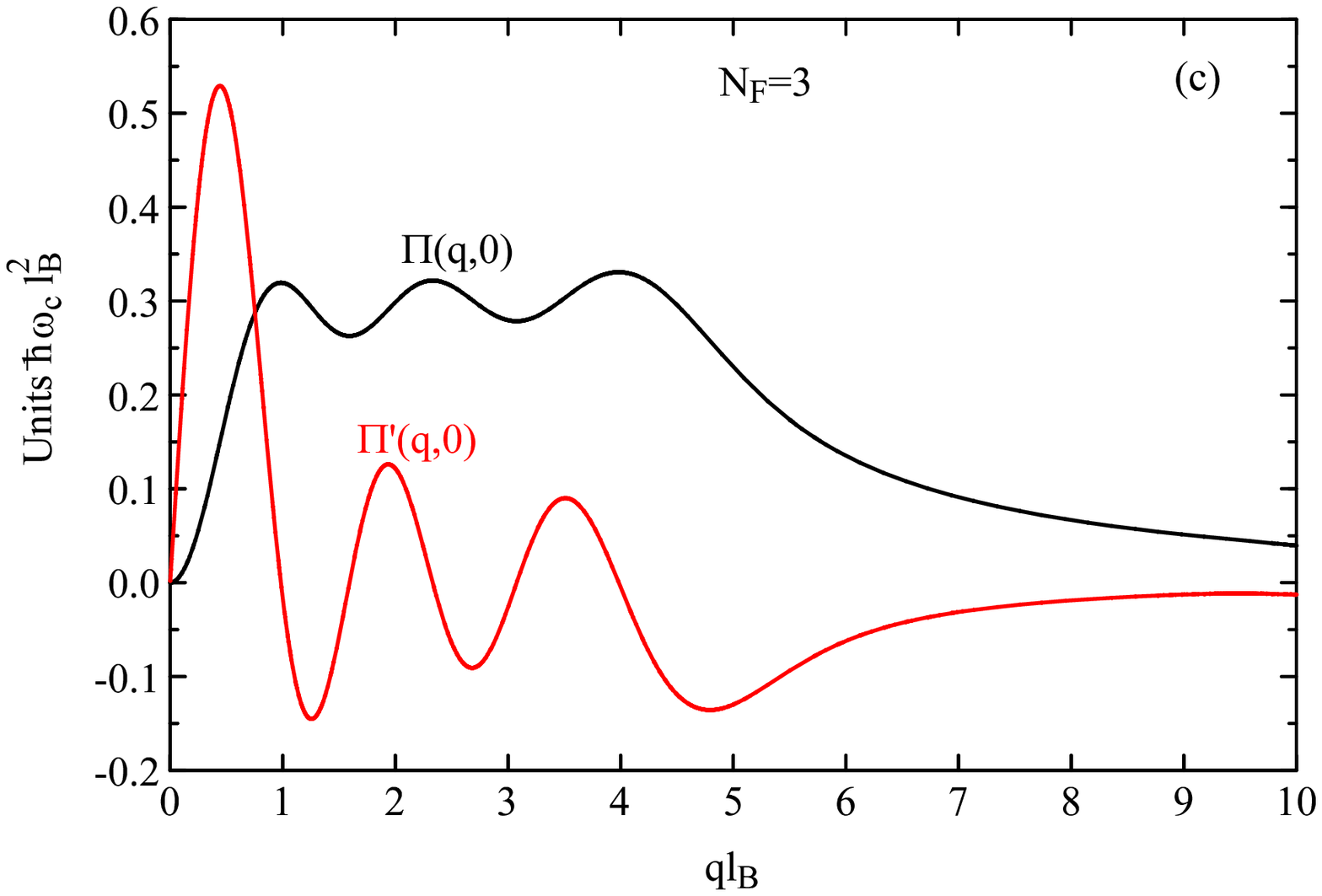}
\includegraphics[width=3.67in,height=2.41in,keepaspectratio]{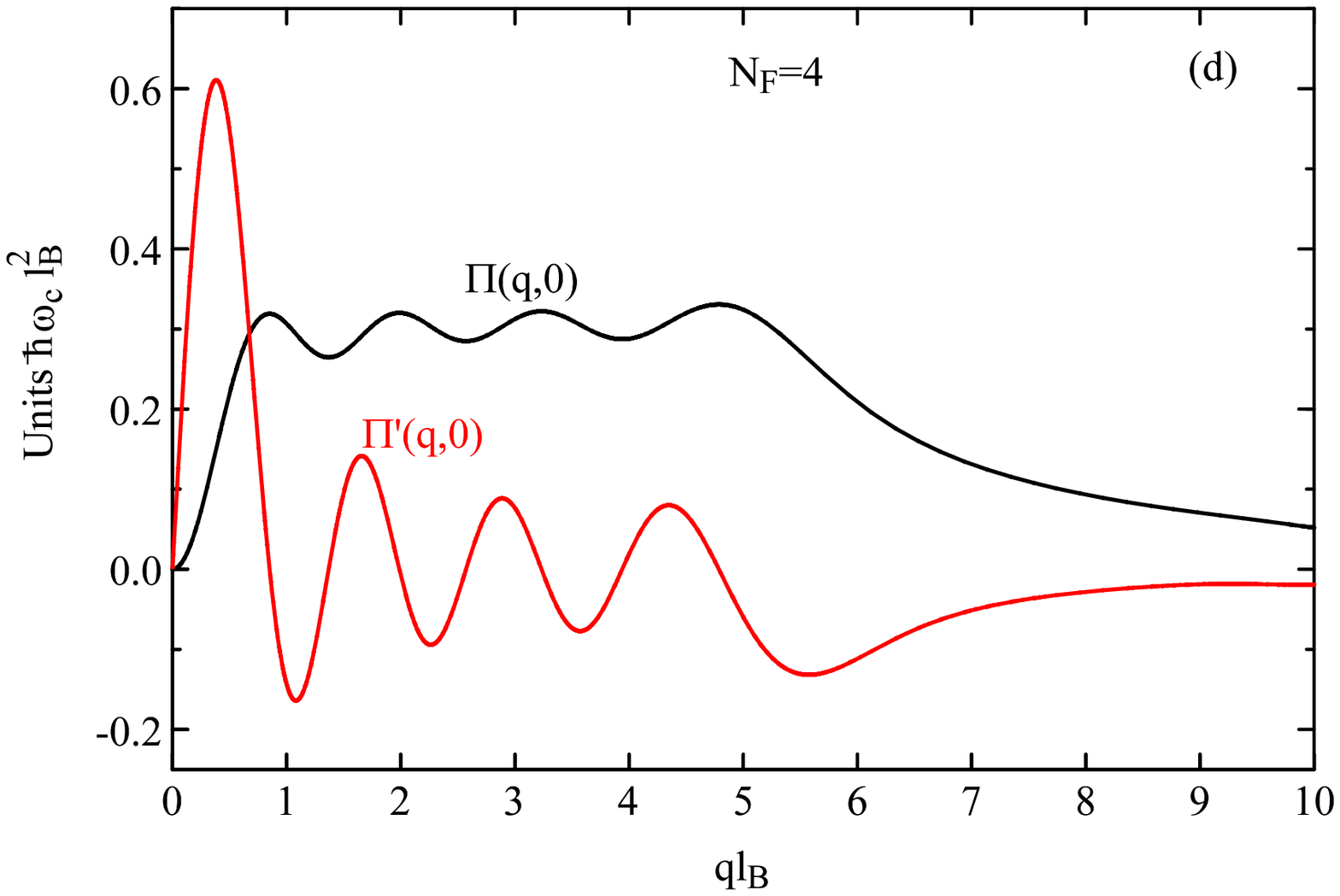}
\caption{(Color online) Static polarization function and its derivative
for the 2DEL at $T=0$ for various filling factors   as a
function of the wave vector $q   $ in units of the inverse magnetic
length $l_B^{-1}$.   }
\label{FIG:2}
\end{figure}

\begin{figure}[t]
\centering
\includegraphics[width=3.67in,height=2.41in,keepaspectratio]{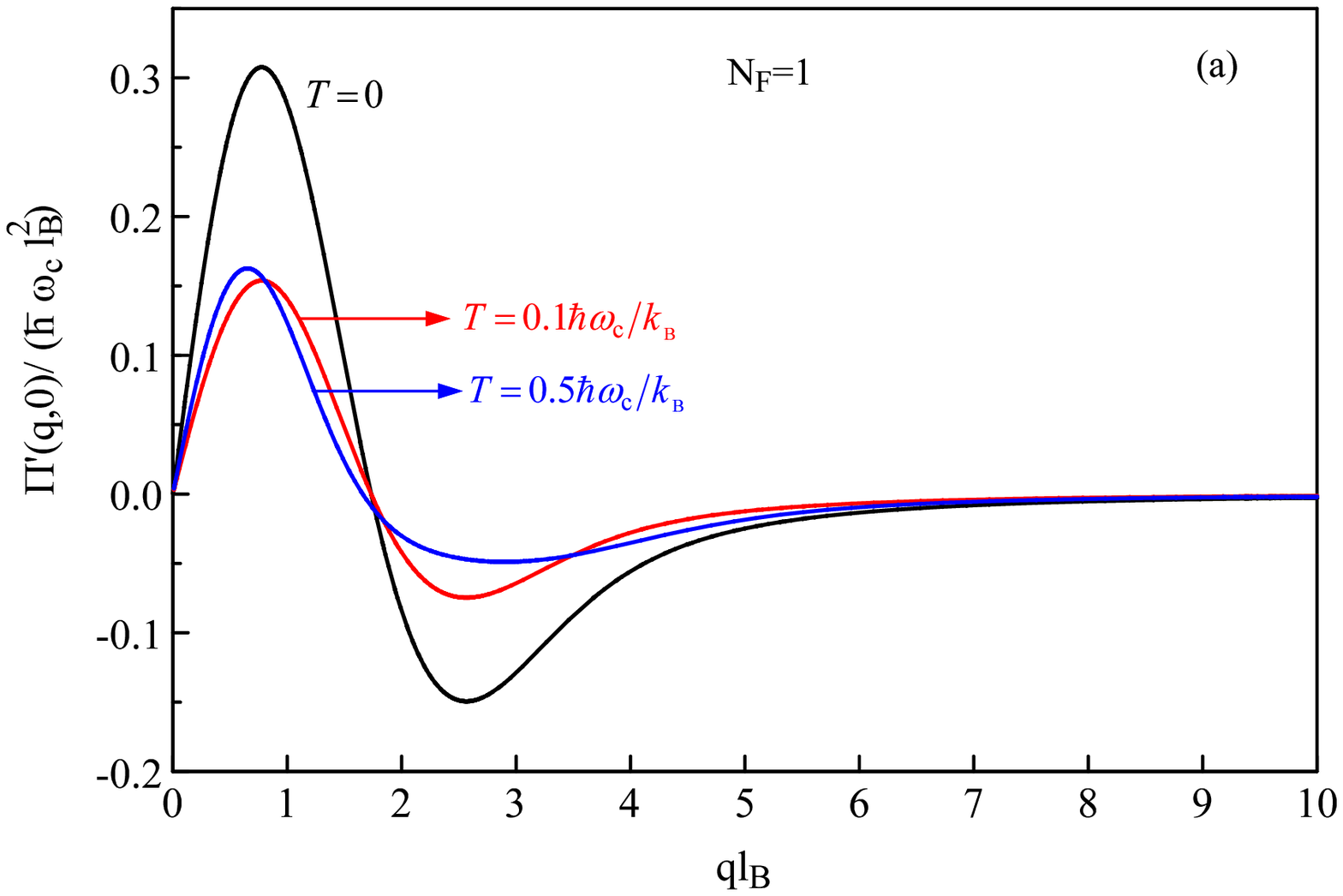}
\includegraphics[width=3.67in,height=2.41in,keepaspectratio]{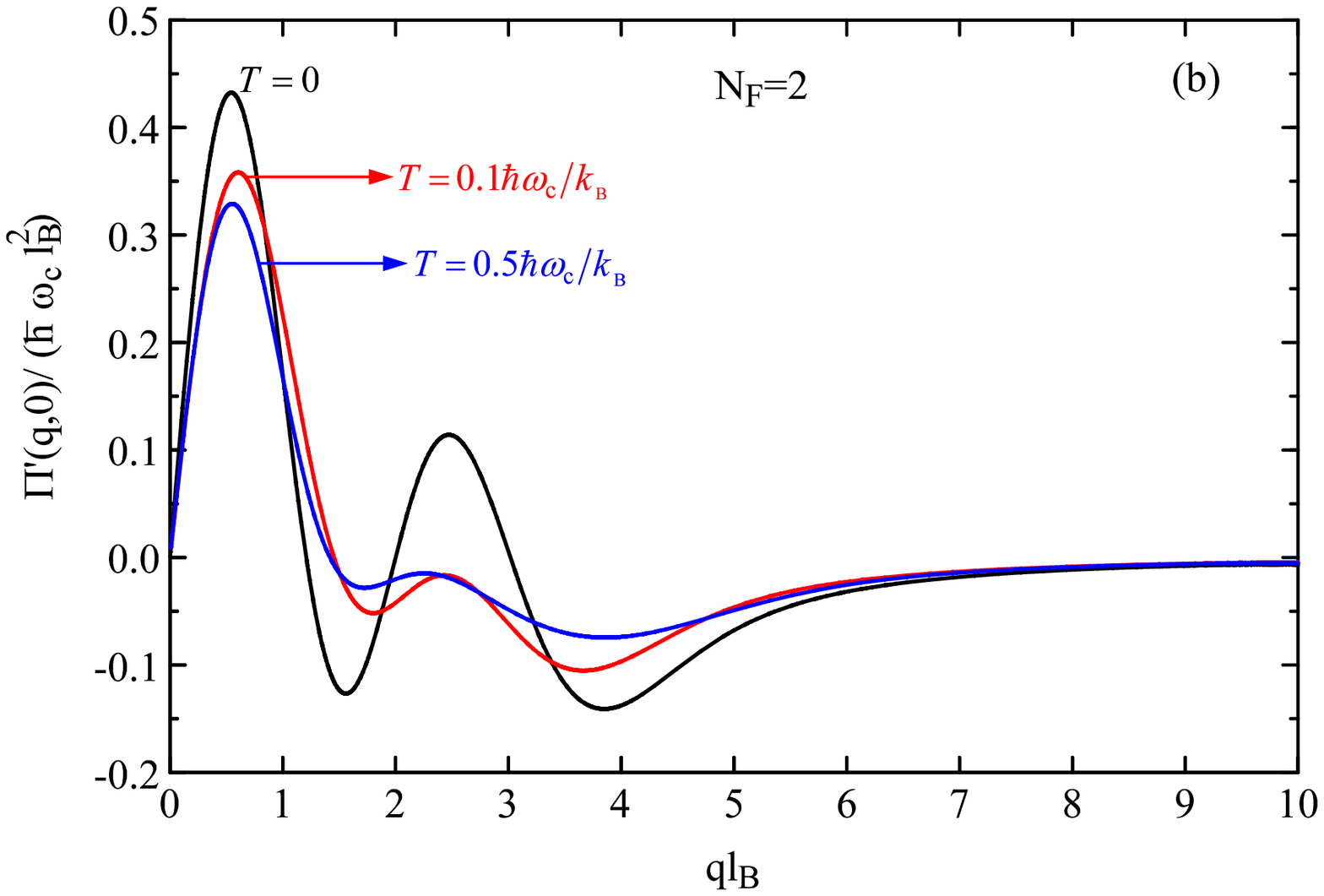}
\caption{(Color online) Derivative of the static polarization function
for the 2DEL at  various temperatures and a chosen filling factor    as a
function of the wave vector $q   $ in units of the inverse magnetic
length $l_B^{-1}$.   }
\label{FIG:3}
\end{figure}

In Fig.\ \ref{FIG:2}, we have presented results for the polarization function and its derivative
for a 2DEL for several filling factors  ($N_F=1,2,3,4$) at $T=0$ K.   Additionally, in Fig.\ \ref{FIG:3},
we have demonstrated what effect temperature has on the derivative of polarization in Fig.\ \ref{FIG:2} for the two cases when
$N_F=1,2$.  These results should be compared with those in Ref.\ \cite{LaryG1} which show that
the part of the curve for the derivative of the polarizability which connects the first local maximum with the
first local minimum is a manifestation of the smoothing of the Fermi surface by temperature.

\medskip
\par

\begin{figure}[t]
\centering
\includegraphics[width=3.67in,height=2.41in,keepaspectratio]{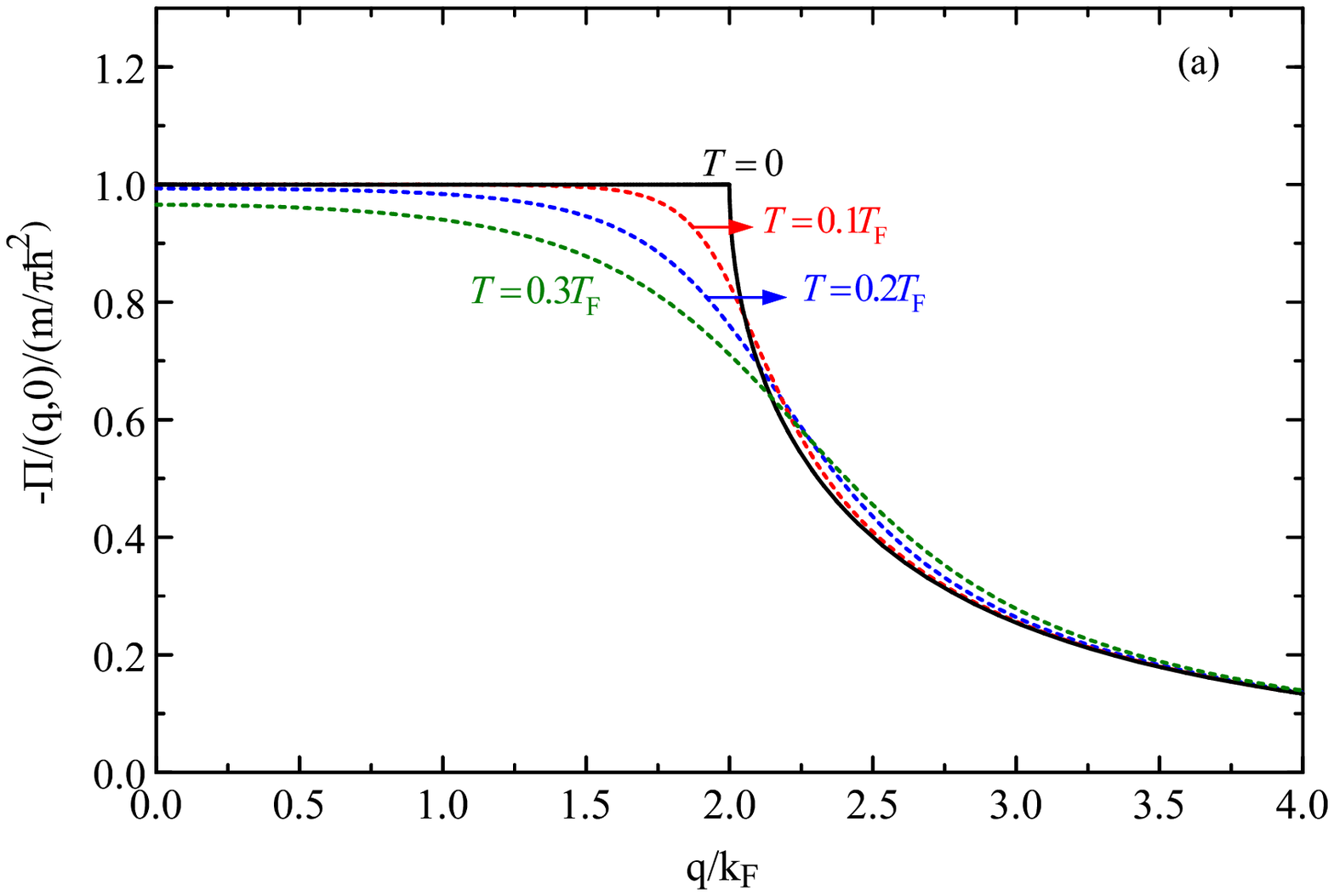}
\includegraphics[width=3.67in,height=2.41in,keepaspectratio]{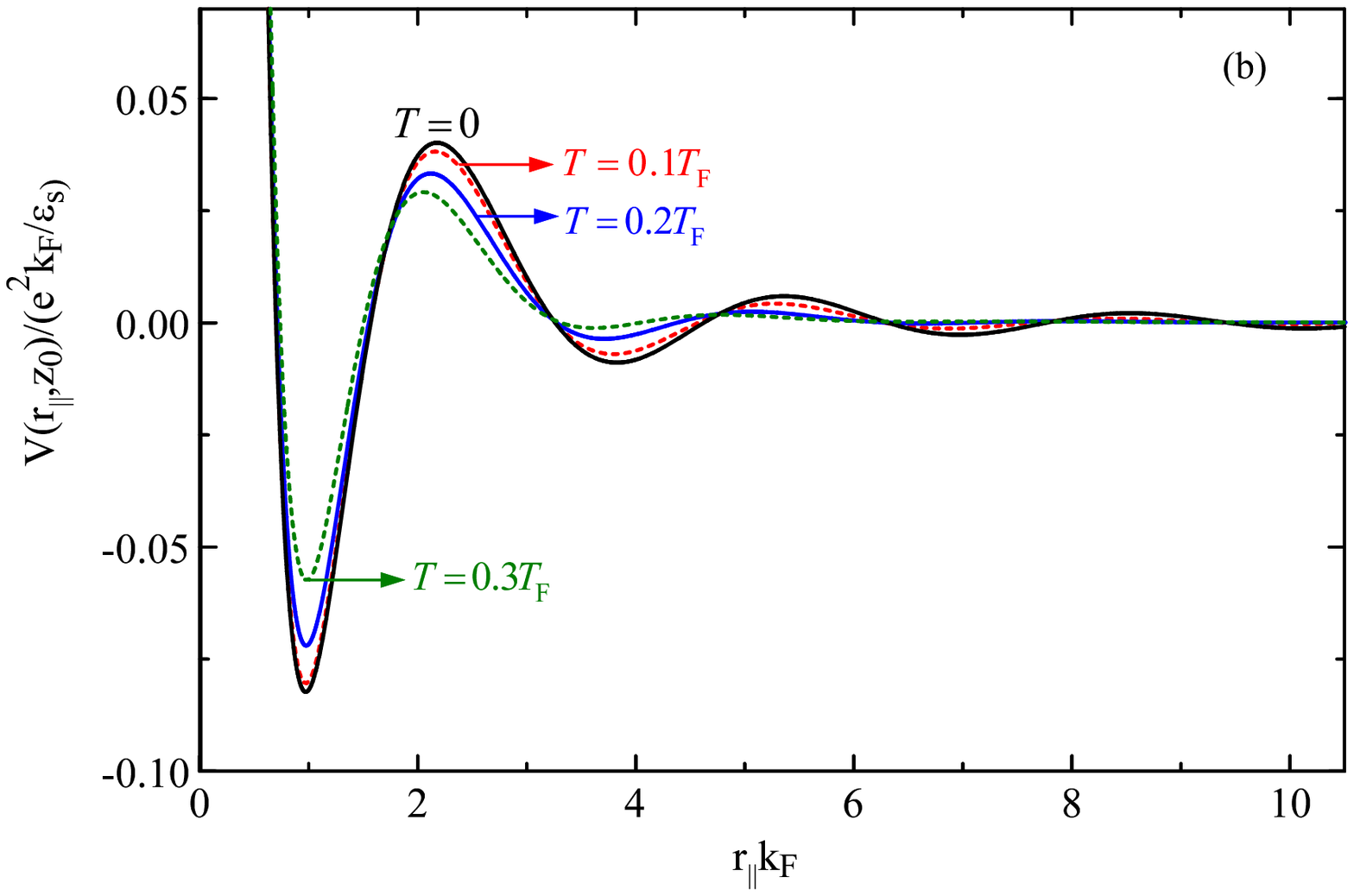}
\caption{(Color online) For various temperatures, the static polarization function   (left panel)
for the 2DEL at   finite temperature in the absence of magnetic field   as a
function of the wave vector $q   $ in units of $k_F$.   The right panel shows
 the corresponding screened  potentials.
 }
\label{FIG:4}
\end{figure}

In the absence of any magnetic field, we may obtain the polarizability at $T=0$ K   by making  use
of the analytical formula of Stern \cite{Stern}. In this case, the derivative  function has a
discontinuity at $2k_F$, i.e., twice the Fermi wave vector.  In fact, between $0<q <2k_F$, the
polarizability is constant and for $q>2k_F$, it falls off like $1/q $.
Although the polarizability function itself is continuous at $2k_F$, the abrupt change in slope
is what gives the discontinuity in its derivative. The plots in Fig.\ \ref{FIG:4}
show that when we introduce temperature, there is an effect on the   polarization
function and subsequently on the screened potential. In accordance with our results above, we
obtain a finite number of oscillatory peaks for a finite magnetic field which  is increased as the
magnetic field is reduced. Therefore, we have, as expected, the Friedel oscillations at $B=0$
in Fig.\ \ref{FIG:4} and whose amplitudes decrease with temperature.

\medskip
\par

At this point, we note that when  we introduce  magnetic field into the calculations,
some distinct features arise. First, each of the plots in  Figs.\ \ref{FIG:1} through
\ref{FIG:3}  for $\Pi^{(0)}(q ,\omega=0)$ for the 2DEL
is   dependent on  wavevector in the long wavelength limit. However the polarization is constant over a wide range ($0\leq q \leq 2k_F$) in the absence of magnetic field. The length scales in our plots for zero and finite magnetic field are related through the equation $k_F l_B=\sqrt {2N_F-1}$. In the shorter wavelengths limit,  the $1/q $  dependence now has  oscillations, which is
 due to the presence of Laguerre polynomials in the  form factor.   Our
conclusion from these calculations  is that the number of peaks in the polarization
function and its derivative is equal to the filling factor $N_F$ for the number of filled
Landau levels. This is also the same as the number of local maxima in the
screened potential in the presence of magnetic field.

\medskip
\par

\begin{figure}[t]
\centering
\includegraphics[width=3.67in,height=2.41in,keepaspectratio]{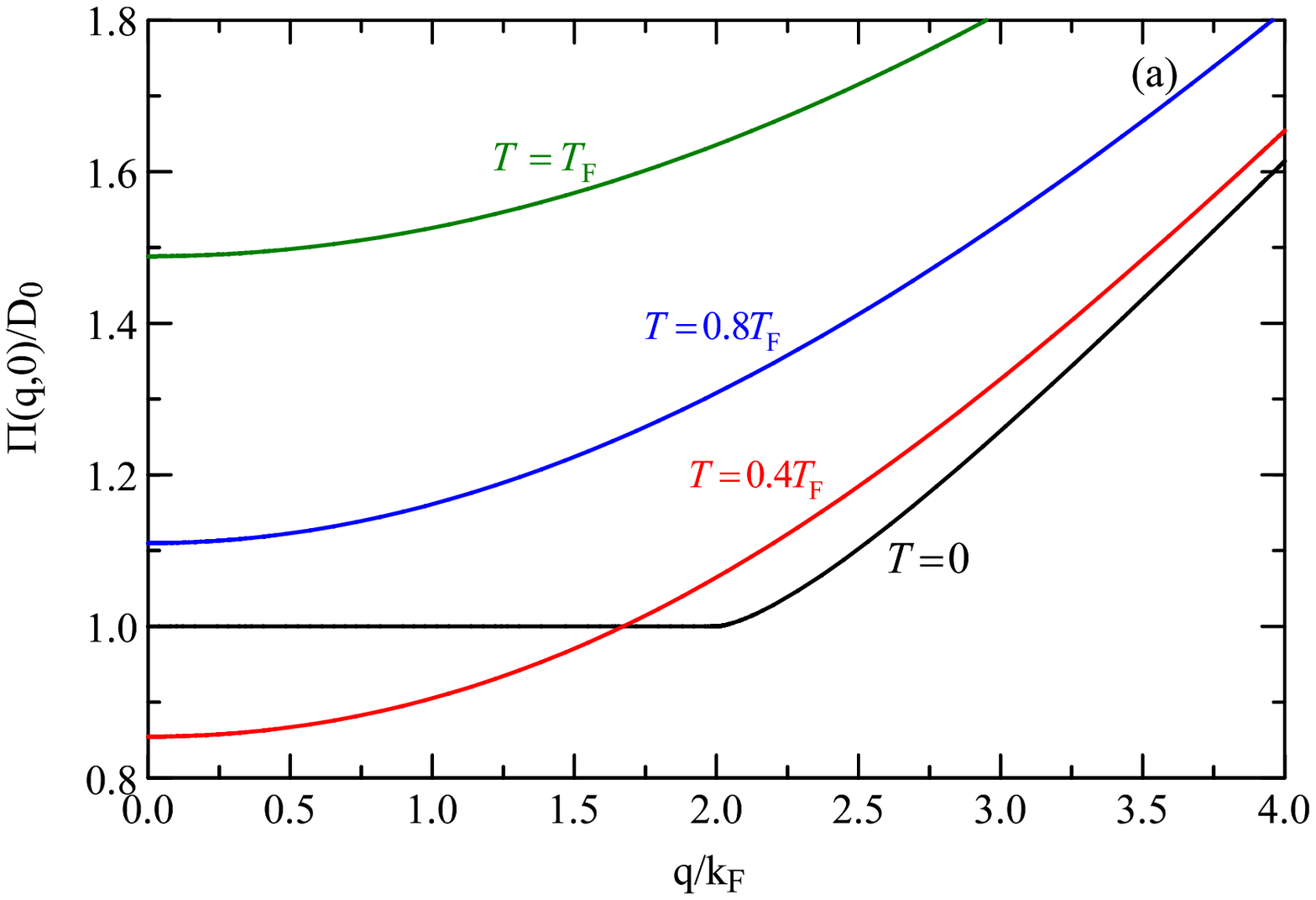}
\includegraphics[width=3.67in,height=2.41in,keepaspectratio]{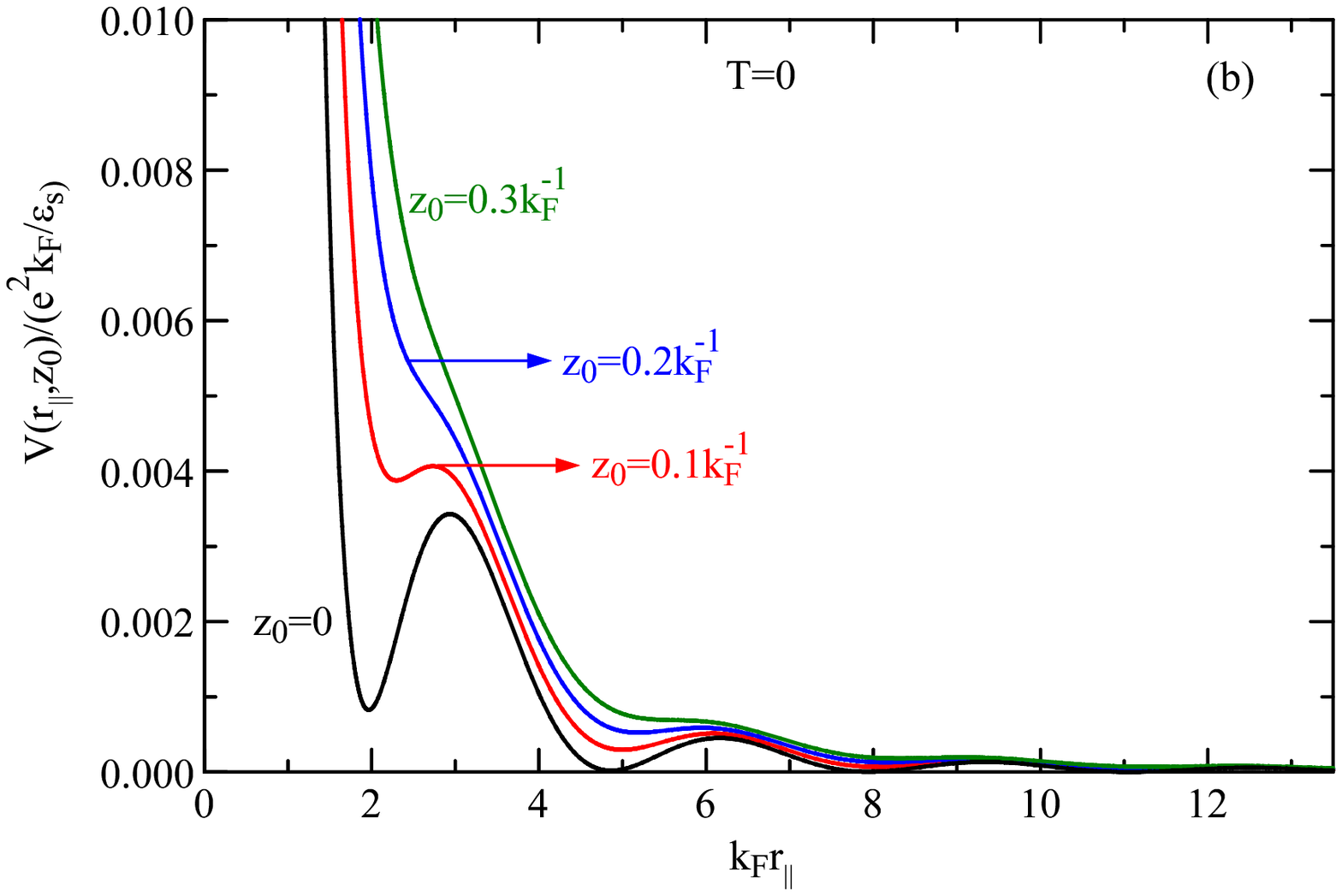}
\includegraphics[width=3.67in,height=2.41in,keepaspectratio]{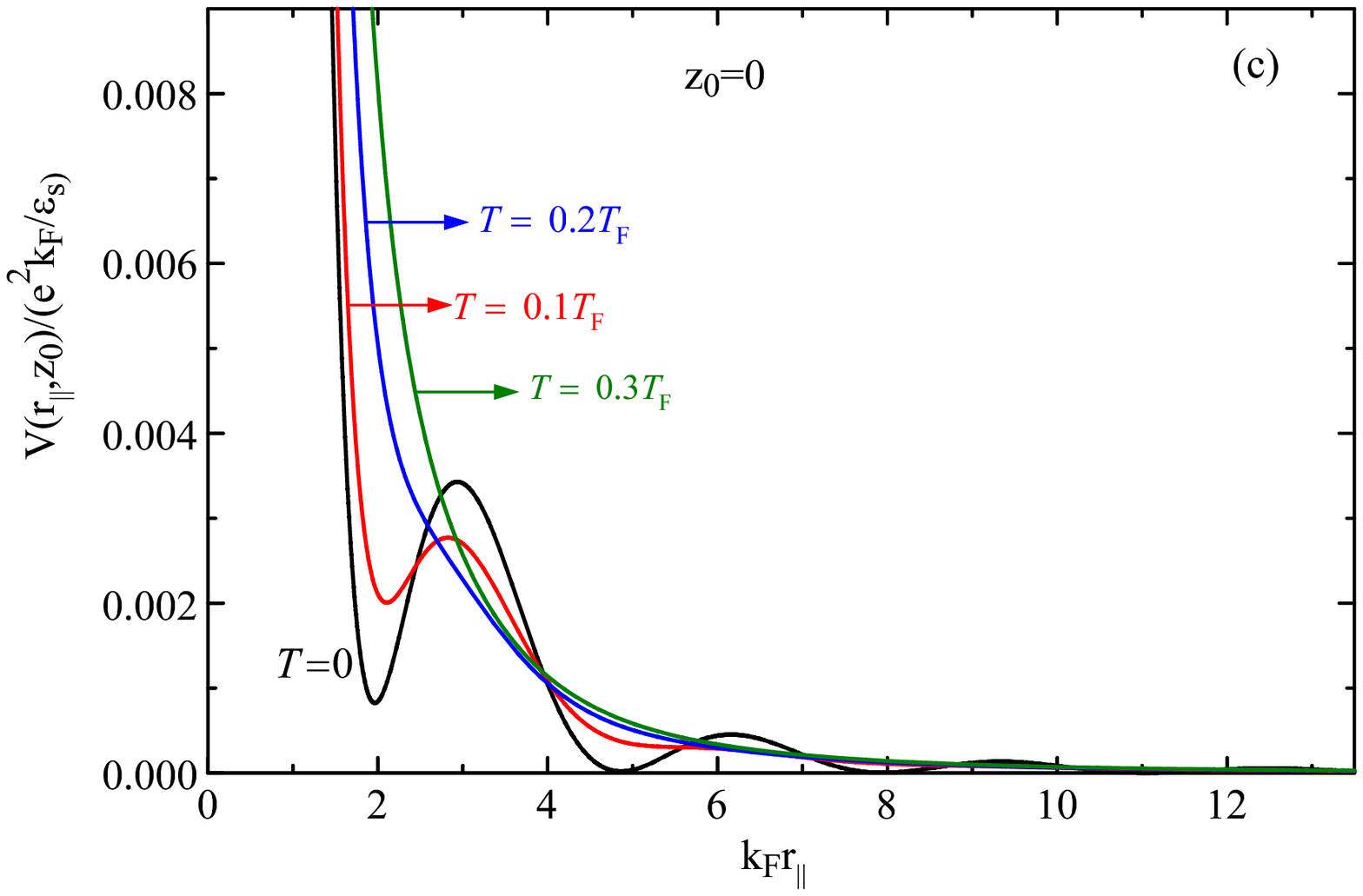}
\caption{(Color online) For Wigner-Seitz-Seitz radius $r_s=1.0$ defined as
$r_s=(e^2/\epsilon_b \hbar v_F) $, the static polarization function  in (a) is plotted
in units of the density-of-states at the Fermi level $D_0=\rho(\epsilon_F)$ for MLG
at $T=0$ as well as finite temperature in the absence of magnetic field   as a
function of the wave vector $q   $ in units of $k_F$.   The   panels (b) and (c)  show
 the corresponding screened  potentials.
 }
\label{FIG:5}
\end{figure}

Turning now to the effect of temperature and magnetic field on {\em doped\/}  monolayer
graphene, we present plots in Fig.\     \ref{FIG:5}  for the   static polarizability and the
screened potential at various temperatures in the absence of an external  magnetic field.
In our numerical simulations, we employ an appropriate value for the Wigner-Seitz
radius defined as the ratio of the potential to the kinetic energy in  an interacting
Coulomb system \cite{MAG5-1}.  For comparison and explanation of the characteristic
features we have obtained, we  refer to the results previously obtained  for  the
static polarization     when $T=0$ K and $B=0$ \cite{wunsch}
 and for finite temperature in the absence of magnetic field \cite{MAG5-2}.
 The effect due to temperature between
$0< q  <2k_F$  is just as significant for doped graphene as for 2DEL since the constant
polarization function in this  range is   independent on $q $ at $T=0$ K but is
definitely wave vector dependent when the sample is heated. Our results in Fig.\ \ref{FIG:5}
for the screened potential show that there may be Friedel oscillations which become
smeared with temperature and also when the charged impurity is moved away from the 2D layer
of graphene. However, the major difference between the Friedel oscillations  for
 the 2DEL and monolayer graphene is that in the former case, there are plus-to-minus
 oscillations as the distance from the impurity is increased. However, for monolayer graphene,
 the screened potential is always repulsive. This, of course, may directly be attributed
 to the difference in band structure for the two media and will indicate the difference
 in the many-body effects due to dielectric shielding.

\medskip
\par

\begin{figure}[t]
\centering
\includegraphics[width=3.67in,height=2.41in,keepaspectratio]{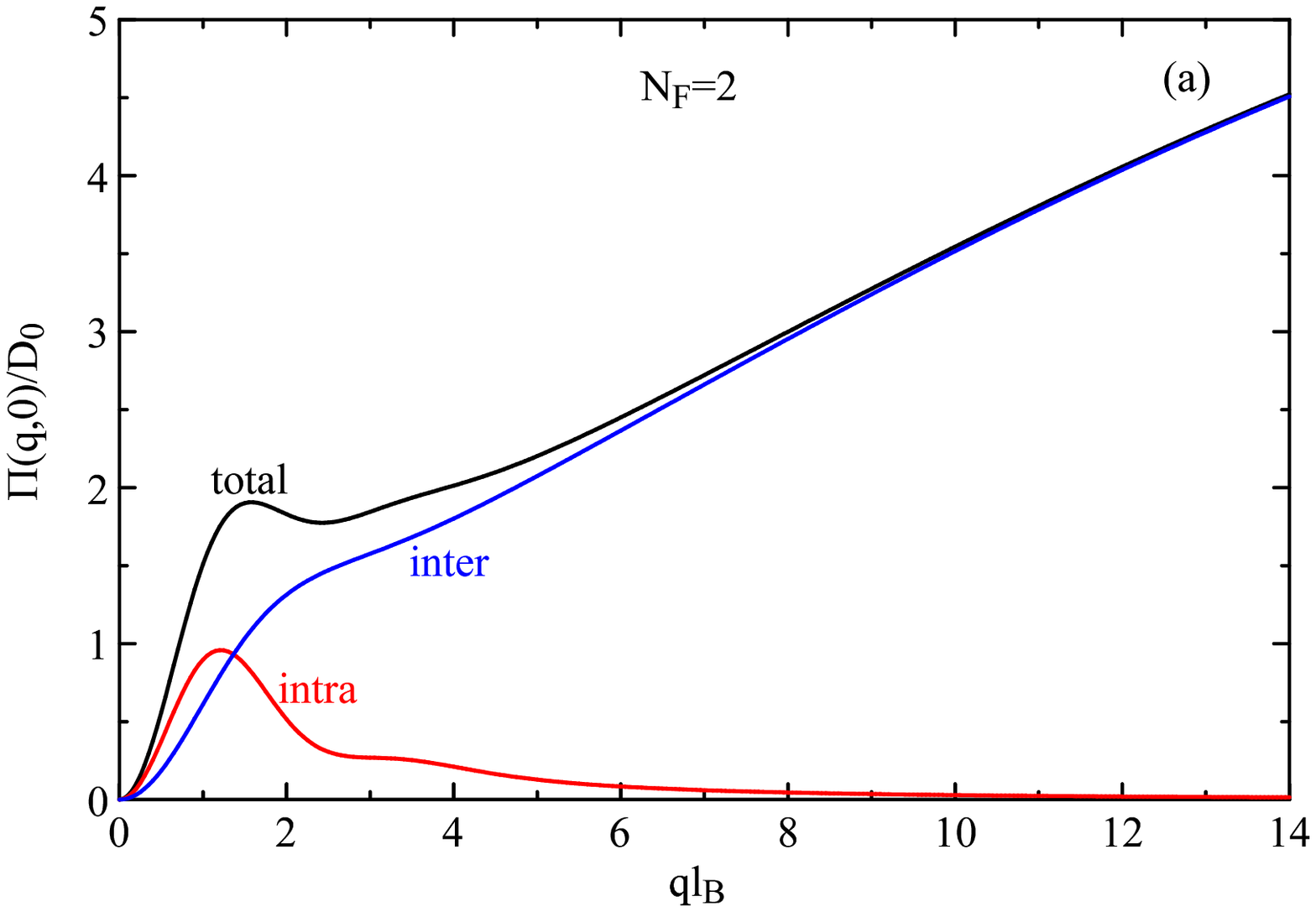}
\includegraphics[width=3.67in,height=2.41in,keepaspectratio]{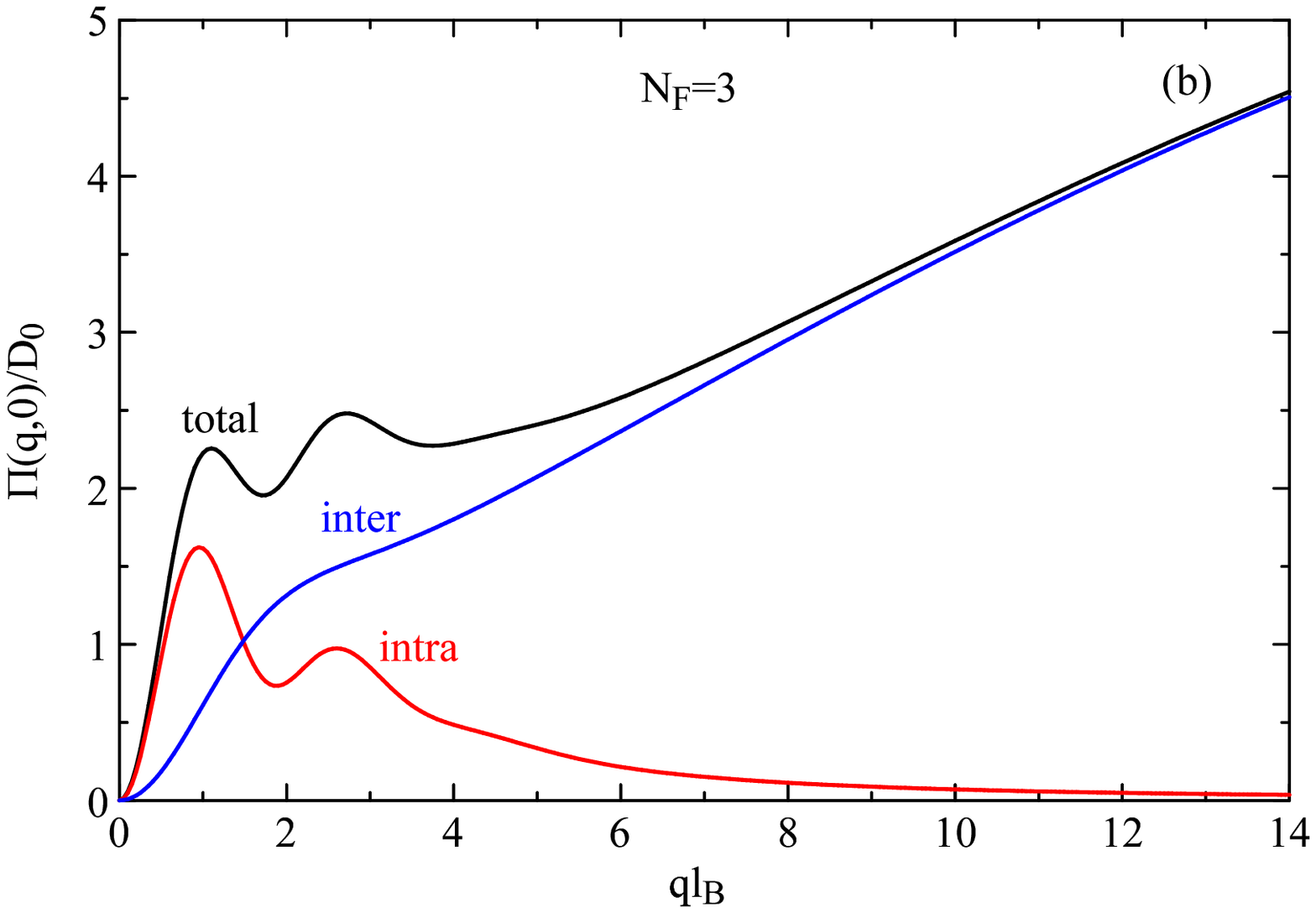}
\caption{(Color online) Static polarization function  for MLG at $T=0$ K  in the  presence of magnetic field
 as a function of the wave vector $q   $ in units of $l_B^{-1}$.   The number of filled Landau levels was
chosen as  (a) $N_F= 2 $ and (b) $N_F= 3 $.  The intra and inter-subband   contributions are presented.
The inset in (b) is for a larger range of wave vector and demonstrates that the polarizability
eventually tends to zero in the short wavelength limit.}
\label{FIG:6}
\end{figure}

Figure \ref{FIG:6} shows our results for the static polarization function
for monolayer graphene at $T=0$ K when there is a magnetic field present.
We present results corresponding to the cases when there are $N_F= 2,3$
Landau levels filled.  For clarity, we separated the contributions
to the polarizability corresponding to intra-band and inter-band transitions.
The intra-band static polarization function has the same number of peaks
as the total number of occupied Landau levels, as is the case for the 2DEL
where only conducting electrons are considered for transitions from
below to above the Fermi level.  For the  inter-band polarization function,
there is no restriction on the number of allowed transitions. Consequently,
this is devoid of the ``signature" peaks as its intra-band counterpart possesses.
Adding these two parts leads us to our presented results which agree with those
given by Roldan, et al.  \cite{MAG1+}. We  emphasize that the Dirac cone approximation (in zero magnetic field) is valid only for a limited range range of energy (0$<\epsilon<$300 meV), thereby requiring that  in
carrying out our calculations, we must introduce   a cut-off value  $N_c$ for the
number of Landau levels in the presence of a uniform perpendicular magnetic field.
Specifically, in Fig.\ 6, we chose the cut-off value $N_c=350$.

\medskip
\par

\begin{figure}[t]
\centering
 \includegraphics[width=3.67in,height=2.41in,keepaspectratio]{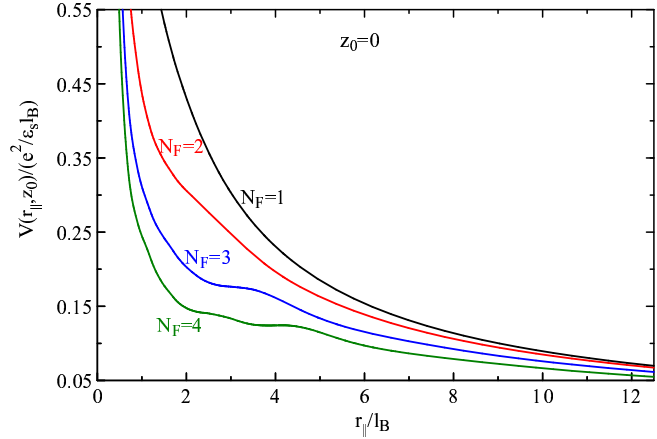}
 \caption{(Color online) The screened potential for monolayer graphene at $T=0 $ K
 in the presence of magnetic field for various $N_F $. The results were obtained for
 chosen  $z_0=0$, i.e., for   an impurity imbedded within  the 2D layer. }
\label{FIG:7}
\end{figure}

Our results above show that for 2DEL, there are $N_F$ peaks in the polarization function
as well as $N_F$ peaks for the screened potential for chosen magnetic field.
For graphene, there are also $N_F$ peaks in the polarization function,
coming from the $N_F$ filled Landau levels, as we noted above. However,  as we show in
Fig.\ \ref{FIG:7},    there are  $N_F$  peaks in the screened potential which are clearly visible but
these are very sensitive to whether   or not the impurity is embedded within the 2D layer. When we set the impurity
at a finite distance from the graphene layer, the peaks in the screened potential are smeared out.
The sensitive nature of  these results  establishes a dramatic difference between MLG and the 2DEL.

\medskip
\par

The results here  not only  provide additional insight into the behavior of the static
polarization function  in strong magnetic fields over a range of  temperature, but they 
demonstrate the many-body effects on the screening properties under these varying environmental
conditions.  Although the emphasis has been on integer filling of the Landau levels,  we may
extend our calculations to the case  when  there is fractional filling. Our calculations show that substantial differences
occur between the 2DEL and monolayer graphene  in the $q\to 0$ limit where the filling factor 
response functions have different functional dependences on the wave vector due to  the 
absence of low-energy excitations when $B\neq 0$, contrary to  the contribution from the energy
bands to the Lindhard function in the absence of magnetic field.  From a mathematical
point of view, this difference is due to the sum over $n,n^\prime$-sum being dominated by the
$n_>-n_<=1$  term in Eq.  (\ref{ref:5}), i.e., by the  highest occupied  and the lowest unoccupied
Landau level.  Figures \ref{FIG:1} through   \ref{FIG:3}   show oscillations for finite filling factor,
for intermediate wave vector. However,  these oscillations  in $\Pi(q,0)$ subside with increasing $N_F$
in accordance with Fig. \ref{FIG:4}, and the screened potential  exhibits the Friedel-like  
oscillations as a function of displacement. Similar conclusions are obtained  for monolayer
graphene, as we have shown.

\section{Concluding Remarks and Summary}
\label{sec4}

We have presented a   comprehensive  description of the combined effect of temperature
and magnetic field on the static polarization function for MLG associated with the Dirac point.
For completeness, we calculated numerically the polarizability for the 2DEL under
a uniform perpendicular magnetic field at finite temperature.  Our results for the 2DEL
are concerned with the effect due to a strong  magnetic field on the  polarizability 
and screening in monolayer graphene and the 2DEL.  This is in contrast with  
 Ref.\ \onlinecite{LaryG1} where the weak field limit was emphasized.    The 
results we obtained are given in an experimentally
achievable long wavelength regime.   Besides,  the numerical results were obtained at
low temperature where the sharpness of the Fermi surface is evident and at high enough temperature
which brought to bear the effect which heating may have on this  property of the electron liquid.

\medskip
\par

Among our main conclusions which arise from  our calculations in this paper, we
have obtained the following.  For doped MLG,   the most important effect occurs in the limit
$q  \to 0$ for any filling factor $N_F$ and any temperature, as well as at
larger wave vectors (beyond $q  >2k_F$) when the temperature is high. At $T=0$ K, it is clear from a mathematical point of view that the polarizability
is dominated by transitions from the highest occupied to the lowest unoccupied Landau level.
As the temperature is increased, other transitions contribute thereby leading to a smearing
of the low-temperature structure.
If the magnetic field is sufficiently high, then  the Landau level separation
is large. Consequently, we  may take only a few terms in the polarization sums since
transitions to higher levels have smaller oscillator strengths.


\newpage

\end{document}